\documentclass{article}
\usepackage[utf8]{inputenc}
\usepackage{fullpage}

\usepackage{graphicx}
\usepackage{amsmath}
\usepackage{amsfonts,amssymb,amsthm}
\usepackage{setspace}
\usepackage[superscript,biblabel]{cite}

\usepackage{booktabs}

\providecommand{\keywords}[1]
{
  \small	
  \textbf{\textit{Keywords---}} #1
}

\DeclareMathOperator{\expit}{expit}
\DeclareMathOperator{\logit}{logit}

\newcommand\independent{\protect\mathpalette{\protect\independenT}{\perp}}
\def\independenT#1#2{\mathrel{\rlap{$#1#2$}\mkern2mu{#1#2}}}

\title{Identifiability and estimation under the test-negative design with population controls with the goal of identifying risk and preventive factors for SARS-CoV-2 infection}

\author{Mireille E. Schnitzer$^{1,2,3,*}$\and Daphna Harel$^{4,5}$ \and Vikki Ho$^{2,6}$ \and Anita Koushik$^{2,6}$ \and Joanna Merckx$^3$}
\date{$^1$Faculty of Pharmacy, Universit\'e de Montr\'eal, Montr\'eal, Qu\'ebec, Canada\\%
    $^2$Department of Social and Preventive Medicine, School of Public Health, Universit\'e de Montr\'eal, Montr\'eal, Qu\'ebec, Canada\\
    $^3$Department of Epidemiology, Biostatistics and Occupational Health, McGill University, Montr\'eal, Qu\'ebec, Canada\\
    $^4$Department of Applied Statistics, Social Science, and Humanities, Steinhardt School of Culture Education and Human Development, New York University, New York, New York, USA\\
    $^5$Center for Practice and Research at the Intersection of Information, Society, and Methodology (PRIISM), Steinhardt School of Culture Education and Human Development, New York University, New York, New York, USA\\
    $^6$Universit\'e de Montr\'eal Hospital Research Centre (CRCHUM), Montr\'eal, Qu\'ebec, Canada.\\
   $^*$ mireille.schnitzer@umontreal.ca \\[2ex]%
    \today}

\begin{document}

\doublespacing

\maketitle

\begin{abstract} 
   Due to the rapidly evolving COVID-19 pandemic caused by the SARS-CoV-2 virus, quick public health investigations of the relationships between behaviours and infection risk are essential. Recently the test-negative design was proposed to recruit and survey participants who are symptomatic and being tested for SARS-CoV-2 infection with the goal of evaluating associations between the survey responses (including behaviours and environment) and testing positive on the test. It was also proposed to recruit additional controls who are part of the general population as a baseline comparison group in order to evaluate risk factors specific to SARS-CoV-2 infection.  In this study, we consider an alternative design where we recruit among all individuals, symptomatic and asymptomatic, being tested for the virus in addition to population controls. We define a regression parameter related to a prospective risk factor analysis and investigate its identifiability under the two study designs. We review the difference between the prospective risk factor parameter and the parameter targeted in the typical test-negative design where only symptomatic and tested people are recruited.
   Using missing data directed acyclic graphs we provide conditions and required data collection under which identifiability of the prospective risk factor parameter is possible and compare the benefits and limitations of the alternative study designs and target parameters. We propose a novel inverse probability weighting estimator and demonstrate the performance of this estimator through simulation study.  \end{abstract}
    


\section{Introduction}
Under the current pandemic caused by the SARS-CoV-2 virus, where the resulting illness is referred to as COVID-19, it is challenging to implement fast epidemiological inquiries to map and understand the disease. Highly infectious~\cite{heinfectious} in a completely non-immune population and targeting primarily the respiratory system with clinical symptoms that include fever, cough, and fatigue~\cite{cummings,wangcovid}, this illness continues to cause substantial morbidity and mortality, straining the healthcare systems of many countries. With the aim of reducing infection, global campaigns encourage individuals to modify their daily behaviour by measures which include physical distancing, the use of masks, and intensified hygienic practices. Much research interest lies in identifying and quantifying any modifiable factors and interventions that may be effective for reducing infection probabilities at an individual or population level. 

Given the challenges involved in testing large portions of the population for active infection with SARS-CoV-2, cases in the general public are typically ascertained through testing sites. In Canada, though regulations have varied by epidemic stage and jurisdiction~\cite{RVIWG}, in order to obtain a test, individuals may be required to be experiencing symptoms of COVID-19 and/or have other reasons to believe that they are infected, such as being a healthcare worker or having recently traveled. Thus, if potential study participants are recruited at test centers, the resulting study sample will not be representative of the general population at risk for the disease. Further, due to the nature of testing self-selection, simple associations measured between participant covariates (e.g. demographics, characteristics, and behaviours) and infection among those tested will not necessarily be representative of true causes or even predictors of infection~\cite{Hernanselection}.

The test-negative design (TND), best known for its use in vaccine effectiveness research, involves the recruitment of symptomatic individuals who are being tested for a given infection and the subsequent comparison between those who tested positive versus negative.~\cite{deserres,fukushima} The fundamental premise of the TND is that all tested individuals are comparable in terms of their symptoms and other reasons for being tested, but may differ in terms of the outcome of being infected by the specific disease of interest,~\cite{Foppa2016,Sullivan} thus making possible comparisons of risk factors between the disease of interest and other related diseases.~\cite{vdb2} 
However, such comparisons do not address the relevant question in the COVID-19 setting where we are interested in which behaviours may increase the risk of becoming infected with SARS-CoV-2 specifically, rather than in relation to another illness or condition leading to the same symptoms.~\cite{vdb2020} 

With the goal of identifying risk/preventive factors specific to SARS-CoV-2, Vandenbroucke et al\cite{vdb2020} proposed a modified TND that combines the recruitment of symptomatic individuals being tested for SARS-CoV-2 infection,  with additional population controls. 
This case-cohort approach compares differences in study participant covariates between people who test positive, people who test negative, and people who are not being tested in order to triangulate factors that likely increase or decrease the odds of SARS-CoV-2 infection. Karmakar and Small\cite{Karmakar} proposed more efficient statistical methods to compare factors between the three groups.  Under this design, with some structural assumptions, one can identify which factors increase (or decrease) the specific risk of contracting SARS-CoV-2. 

Several studies~\cite{Sullivan, Shi2017} used directed acyclic graphs (DAGs) to illustrate various potential sources of bias in a TND for the goal of evaluating vaccine effectiveness. While DAGs have been described as limited,~\cite{vdb2} 
 they can helpfully identify data structures where there is a potential for bias that is not corrected by the study design. In particular, Sullivan et al~\cite{Sullivan} demonstrated the potential for selection bias when the TND and analysis do not fully control for  health-care-seeking behaviours.

In order to provide researchers with insight and additional options for the rapid  evaluation of risk factors of SARS-CoV-2, this methodological study investigates an alternative design that recruits all SARS-CoV-2-tested patients (including asymptomatic) in addition to population controls. We evaluate plausible data structures and highlight the data collection necessary to formally allow for unbiased estimation in a ``risk/preventive factor'' analysis corresponding to modeling the covariates that are prospectively predictive of SARS-CoV-2 infection. We also propose a novel estimator that is consistent under the given design and assumptions.

In the following sections, we define our parameter of interest and differentiate our parameter with respect to the target parameter of the TND that only recruits \emph{symptomatic} tested people (hereafter called the ``proper'' TND). We then provide and discuss identifiability conditions of our parameter of interest under our alternative design in the context of the COVID-19 pandemic. 
 Identifiable means that we would know the exact value of the parameters of interest if we had an infinite sample size; our study thus gives some conditions, including requirements for data-collection, under which unbiased estimation of our target parameter can be achieved.  
Finally, we propose a feasible inverse probability weighting (IPW) estimator~\cite{HTest,robinssemipar1995,ColeHernanIPW}  that allows for consistent estimation under the general data-generating structure, without parametric assumptions. Compared to the proper TND, it requires additional data collection but it also allows for the statistical adjustment of measured confounders of infection and testing. The proposed method could be used as an alternative or a complement to the investigative triangulation approach previously proposed.~\cite{vdb2020} Through simulation study, we then contrast various implementations of our estimator under our modified study design with analyses conducted under the proper TND, with and without additional population controls. The discussion and eAppendix Section 2 evaluate and contrast potential sources of bias under the different study designs.

\section{The Test-Negative Study Design with Population Controls}

Given access to the recruitment of people seeking SARS-CoV-2 tests at a given testing site, we consider an alternative study design that involves the recruitment of two groups of people: (1) people who are being tested for SARS-CoV-2 at the test site and (2) members of the general population who are not being tested but who are under the jurisdiction of the test site. It must be the case that members of group (2) would be able to access the test site were they to have symptoms of COVID-19 or otherwise qualify for and seek testing. 
Participants recruited from group (1) are denoted $T=1$ and those from group (2) are denoted $T=0$. This is essentially a case-control study design except that cases are those who are tested and controls are those who are untested. We assume for the current development that members of both groups are independent, with a total sample size of $n$. The need for independence implies that, for example, only one member per household should be recruited.

All participants are given a questionnaire to collect information about the potential risk/preventive factors under study, $\boldsymbol X$, which we will now refer to simply as risk factors, and related confounders, $\boldsymbol C$. The questionnaire may also capture information about current symptoms related (or believed to be related) to COVID-19, $\boldsymbol W$. It is also necessary that we receive the result of the SARS-CoV-2 test from those being tested. 

Recruitment from groups (1) and (2) will give us three categories of participants: those being tested and who test positive for SARS-Cov-2 (``test-positives''), those being tested and who test negative for SARS-Cov-2 (``test-negatives''), and those not being tested. The analysis will incorporate information from the three groups with the aim of identifying risk factors $X$ that may increase or decrease the probability of becoming infected with SARS-CoV-2.

Note that the design described above is distinct from a proper TND where the inclusion criteria requires one or more symptoms characteristic of COVID-19 disease.~\cite{vdb2020}  

\section{Parameters of Interest}\label{sect_param}
In a typical risk factor analysis, the scientific objective is to identify which factors, e.g. behaviours or characteristics of individuals, are prospectively associated with SARS-CoV-2 infection in a chosen outcome regression model, possibly after adjustment for suspected confounders~\cite{Shmueli,Schooling}. We consider the population of interest to be members of the general public who are at risk of infection and who are under the jurisdiction of the SARS-CoV-2 testing site under study. 
The regression model represents the associations between the risk factors and prospective short-term risk of infection with SARS-CoV-2 in this population.

The binary outcome of interest is infection with SARS-CoV-2 (yes/no), denoted $Y^1$. Only those tested have an observed value of $Y^1$. Due to imperfect test sensitivity and specificity, the measured outcome (test result) may not correspond to the true  SARS-CoV-2 infection status but we ignore this complication in the main manuscript.  
The observed data are thus of the form $\boldsymbol O=(\boldsymbol C,\boldsymbol X,T\times Y^1,\boldsymbol W,T)$. The complete data with all outcomes measured are $O^{*}=(\boldsymbol C,\boldsymbol X,Y^1,\boldsymbol W,T)$.  
We will use lower case letters to represent realizations of these random variables. In particular, we observe $\boldsymbol o_i=(\boldsymbol c_i,\boldsymbol x_i,t_i\times y^1_i,\boldsymbol w_i,t_i)$ for individuals $i=1,...,n$ where $n$ is the total sample size.

Then, under the complete data $O^{*}$, we define the logistic regression model
\begin{align}\logit\{\Pr(Y^1=1\mid \boldsymbol X=\boldsymbol x, \boldsymbol C= \boldsymbol c)\} &=  \boldsymbol x^\intercal \boldsymbol\beta + \boldsymbol c^\intercal \gamma \label{mod1}\\
&= \beta_0 + \beta_1 x_1 + \beta_2 x_2 +...+\beta_r x_r + \gamma_1 c_1 + \gamma_2 c_2 ... + \gamma_s c_s \notag
\end{align}
where $\boldsymbol X=(1,X_1,X_2,...,X_r)$, $\boldsymbol C=(C_1,C_2,...,C_s)$ (and similarly for the realizations denoted by lower case letters), $\boldsymbol \beta=(\beta_0,\beta_1,...,\beta_r)$, and $\boldsymbol \gamma = (\gamma_1, \gamma_2,...,\gamma_s)$, under maximum likelihood estimation. Our interest lies in the vector parameter $\boldsymbol \beta$ where $\exp(\beta_k)$ corresponds to the conditional odds ratio (OR) related to the covariate $X_k$.

We note that the parameters in this regression model may not represent causal effects, i.e. even if a coefficient $\beta_k$ is negative, it does not necessarily mean that $X_k$ decreases the risk of infection~\cite{Shmueli,Schooling}. In order to establish such a relationship, all causes of interest must be independently manipulable, all confounders must be adjusted for in the model, and the model must correctly represent the mechanisms of infection. 

Importantly, the parameters of interest we describe here are distinct from those estimated in a proper TND which recruits individuals who are \emph{symptomatic} and being tested. Define $\overline{\boldsymbol W}$, a summary of $\boldsymbol W$, as symptomatic status (yes/no) as defined by the study inclusion criteria.  Under certain conditions~\cite{vdb2}, a logistic regression will allow for unbiased estimation of ORs corresponding to the association of the risk factor with the risk of SARS-CoV-2 conditional on the presence of symptoms that satisfy the inclusion criteria, i.e. $\exp(\boldsymbol\beta^*)$ in
\begin{equation}
\logit\{\Pr(Y^1\mid \boldsymbol X=\boldsymbol x, \boldsymbol C= \boldsymbol c, \overline{\boldsymbol W}=1)\}=  \boldsymbol x^\intercal \boldsymbol\beta^* + \boldsymbol c^\intercal \boldsymbol \gamma^*. \label{mod2}
\end{equation}

Because it is not prospective (i.e. it conditions on symptoms which occur after infection) the parameter $\boldsymbol \beta^*$ has a challenging interpretation; for example, due to differences in infection prevalence, potential for symptoms, and non-collapsibility of the OR,~\cite{Westreich2016} the coefficient of a risk factor may not be null even if the risk factor is exactly as strong on the OR scale for both the infection of interest and other infection. We give a numerical example of this in the simulation study in Section~\ref{sect_simulation}. 

\section{Controlling for Collider Bias Resulting from Selecting Patients at Test Sites}\label{sect_collider}

\begin{figure}[ht]
\centering
\includegraphics[width=\textwidth]{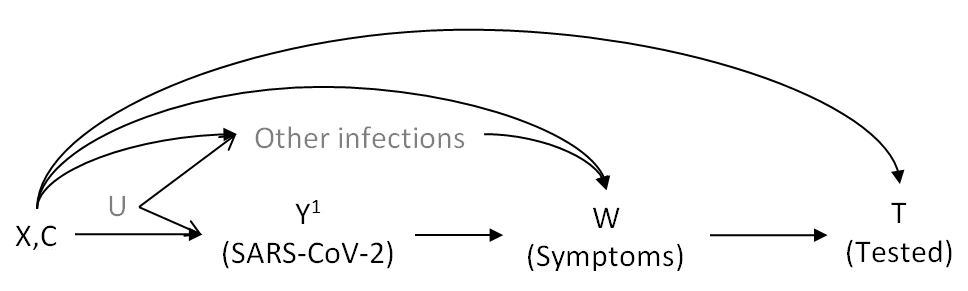}
\caption{mDAG representing hypothetical relationship between baseline covariates $\boldsymbol X$ and $\boldsymbol C$, symptoms $\boldsymbol W$, tested $T$, and infection $Y^1$. Note that $Y^1$ is only observed for tested subjects ($T=1$). The node ``Other infections" is unmeasured, as is $U$. }\label{DAG1}
\end{figure}

The challenge in comparing patients who tested positive versus negative arises from selecting on patients who seek and receive testing. 
Figure~\ref{DAG1} is a missing data directed acyclic graph (mDAG)~\cite{daniel,Mohan} representing assumed relationships between covariates in this analysis. In particular, we allow for the baseline covariates to potentially cause (i.e. influence the risk for) SARS-CoV-2 infection, $Y^1$. If a patient is tested ($T=1$) then we observe a test result $Y^1$; if $T=0$ then $Y^1=NA$. 
The probability of testing depends on whether the individual has suspected symptoms of COVID-19, $\boldsymbol W$ (which may include fever, respiratory symptoms, etc), the risk factors ($\boldsymbol X$), and other baseline covariates ($\boldsymbol C$). For example, an alert individual who frequently hand washes may be more inclined to seek testing, possibly also depending on whether they are experiencing real or perceived symptoms of COVID-19 (included in $\boldsymbol W$). Any variable in $\boldsymbol X$, such as recent travel, that places a person at higher risk for infection may also prompt that person to seek testing, even with absent or mild symptoms.
We assume that infection only affects testing through symptoms. Thus, in $\boldsymbol W$ we include all symptoms known to the participant.  Indeed, $\boldsymbol W$ may reflect symptoms of other  (respiratory or other) disease, and these may be caused by pathogens other than the SARS-CoV-2 virus. We also allow for unmeasured common causes of SARS-CoV-2 and other infections. In constructing this mDAG, we assume that no other unmeasured factors simultaneously affect any pair of nodes. 

The objective described in Section~\ref{sect_param} is to estimate the model parameters representing the relationship between $\boldsymbol X$ and $Y^1$ while adjusting for  $\boldsymbol C$, i.e. the parameters of the model for  $\Pr(Y^1=1\mid \boldsymbol X=\boldsymbol x,\boldsymbol C=\boldsymbol c)$. But because we only have outcome data from those who are seeking testing, we may consider directly modeling the observed outcomes among those who were tested $\Pr(Y=1\mid \boldsymbol X=\boldsymbol x,\boldsymbol C=\boldsymbol c, T=1)$. 
Such modeling of the selected population may produce misleading associations between $\boldsymbol X$ and $Y^1$. This is due to collider bias~\cite{Hernanselection,colecollider}, which is caused by subsetting or adjusting for a variable that is caused by the two variables whose association is of interest. In our case, we would be conditioning on $T=1$, which is caused by both $Y^1$ (through $\boldsymbol W$) and by $\boldsymbol X$. Thus, there is a possibility for erroneous conclusions resulting from the measured associations between $\boldsymbol X$ and $Y$ among those seeking testing.

\begin{figure}[ht]
\centering
\includegraphics[width=\textwidth]{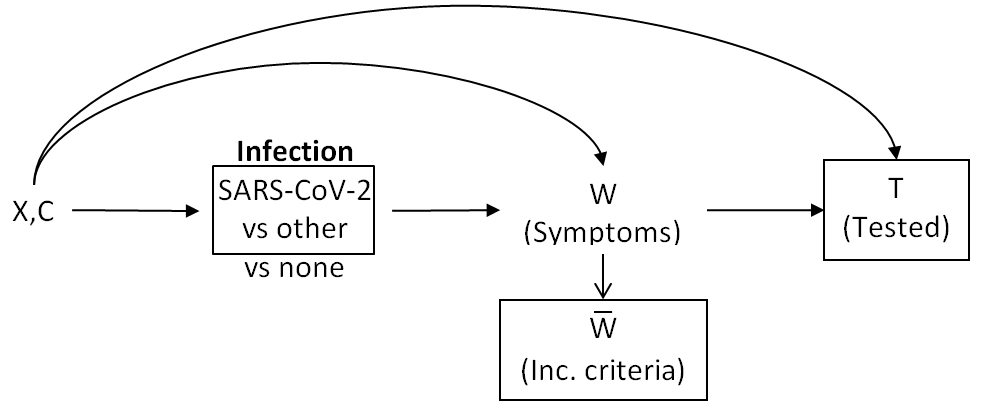}
\caption{mDAG representing hypothetical relationship between baseline covariates $\boldsymbol X$ and $\boldsymbol C$, symptoms $\boldsymbol W$, tested $T$, and infection $Y^1$. The boxes around $T$ and $\overline{\boldsymbol W}$ represents the proper TND which selects those both tested and symptomatic. The box around the two components of the infection variable indicate the fact that, due to selection on symptoms, this design will include only those with COVID-19 or another symptomatic illness.}\label{DAGrel}
\end{figure}

As represented in the mDAG in Figure~\ref{DAGrel}, the proper TND recruits symptomatic patients.  Consequently, an analysis comparing test-positives and -negatives will effectively be conditional on $\overline{\boldsymbol W}=1$. Subsequently, by assumption, this conditioning limits the sample to those diagnosed with COVID-19 and to those who have some other illness resulting in similar symptoms. As in a case-control design, standard results show that a logistic regression analysis will allow for estimation of the OR $\exp(\beta^*)$, which is conditional on $\overline{\boldsymbol W}=1$. By definition, this is biased for the parameter $\exp(\beta)$. If, however, we use a study design that recruits all individuals being tested, $\boldsymbol W$ could alternatively be controlled by statistical adjustment. The benefit of the latter is that, with the addition of population controls, the prospective risk parameter $\boldsymbol \beta$ becomes nonparametrically identifiable. The limitations include the additional recruitment and the requirements that all known symptoms, $\boldsymbol W$, must be collected and assumed to be fully documented. 

To illustrate with an example: access to a private vehicle allows one to avoid public transit, which may be a preventive factor for any viral infection. Thus, we may be interested in measuring the decrease in risk of SARS-CoV-2 related to access to a private car in the general population. However, access to a SARS-CoV-2 test site is also facilitated by access to a vehicle, especially for those living further away from the test sites. Because (we are supposing) those with a car are more likely to be able to seek testing if they have symptoms, there is a disproportionate number of people without cars with symptoms of COVID-19 who are not tested. Thus, under the study design where we recruit among all tested individuals, we may measure a negative association between access to a private vehicle and SARS-CoV-2 even if there is no causal relationship. If we subset our analysis to symptomatic individuals, we will avoid the collider bias, but if the protectiveness of car ownership is similar for COVID-19 and illnesses with similar symptoms, we may obtain an estimate that is close to the null, roughly reflecting the contrast in risk factors between illnesses. But neither of these analyses would address whether car ownership keeps you safe from COVID-19 specifically. This motivates the additional recruitment of population controls.


\section{Nonparametric Identifiability of $\boldsymbol{\beta}$ and Estimation with IPW}\label{ident}

In the eAppendix Section 1, we establish the nonparametric identifiability of the parameters of the model in equation~(\ref{mod1}) under the mDAG in Figure~\ref{DAG1} and the design involving the recruitment among all tested people and among the general population. 


As a consequence of the mDAG in Figure~\ref{DAG1}, we have the independence condition
\begin{align} T \independent{} Y^1 \mid \boldsymbol X=\boldsymbol x, \boldsymbol C=\boldsymbol c, \boldsymbol W=\boldsymbol w \label{indep}
\end{align}
on which the identifiability argument is based. 
Thus, for unbiased estimation we need complete data on all mediators of $Y^1$ and $T$, and mediators of other infections and $T$, which most likely correspond to known symptoms that led to testing.  We must also have measured all common causes of testing and $\boldsymbol W$. As in the proper TND~\cite{Sullivan, Jackson2018}, we must measure all common causes of $X$ and $Y^1$ in order to have interpretable results. Both designs rely on statistical adjustment for common causes of $Y^1$ and $T$, which may arguably include health-care-seeking behaviour~\cite{Sullivan}.  See eAppendix Sections 2.1-2.4 for discussion of the consequences of violations of the mDAG structure. 



One may  construct an estimator using IPW~\cite{ColeHernanIPW,vdl_casecontrol}.
Defining ${Q}_{Y^1,T=1}(\boldsymbol x_i, \boldsymbol c_i, \boldsymbol w_i)=\Pr(Y^1=1\mid \boldsymbol X=\boldsymbol x, \boldsymbol C=\boldsymbol c, \boldsymbol W= \boldsymbol w, T=1)$, the IPW estimator for the $\boldsymbol \beta$ parameter in equation~(\ref{mod1}) is given through the score equations of a weighted logistic regression 
\begin{align}
     \sum_{i=1}^n  \begin{pmatrix} \boldsymbol x_i\\ \boldsymbol c_i \end{pmatrix} \frac{I(t_i=1)\{\hat{Q}_{Y^1,T=1}(\boldsymbol x_i, \boldsymbol c_i, \boldsymbol w_i) - \expit(\boldsymbol x_i \boldsymbol \beta + \boldsymbol c_i \boldsymbol \gamma)\}}{\hat{\Pr}(T=1\mid \boldsymbol X=\boldsymbol x_i, \boldsymbol C=\boldsymbol c_i, \boldsymbol W= \boldsymbol w_i)}=0\label{IPW_eqn}
\end{align}
where values $(\boldsymbol x_i, \boldsymbol c_i, \boldsymbol w_i, t_i)$ refer to the data realizations of subject $i$, $\begin{pmatrix} \boldsymbol x_i\\ \boldsymbol c_i \end{pmatrix}$ represents a column vector, and $\hat{Q}_{Y^1,T=1}(\boldsymbol x_i, \boldsymbol c_i, \boldsymbol w_i)$ represents estimates of ${Q}_{Y^1,T=1}(\boldsymbol x_i, \boldsymbol c_i, \boldsymbol w_i)$. 
 In order to estimate the numerator of the IPW estimator, we must first define a model for ${Q}_{Y^1,T=1}(\boldsymbol x, \boldsymbol c, \boldsymbol w)$. This model is fit on subjects who received a test. 

In order to estimate the denominator of~(\ref{IPW_eqn}), we note that the associations between covariates, symptoms, and the probability of testing must be estimated from the data resulting from the case-control design, where sampling is carried out in both the tested and untested groups. If we know the baseline testing prevalence $q_0=\Pr(T=1)$, we may use a simple weighting method for case-control studies~\cite{vdl_casecontrol}. Specifically, we assign all cases the weight $q_0$ and all controls the weight $(1-q_0)/J$ where $J$ is the ratio of the number of controls to cases in the sample. We use these weights in any chosen binomial regression model for $T$ conditional on $\boldsymbol X$, $\boldsymbol C$, and $\boldsymbol W$. Finally, we use predictions from this model fit to estimate $\hat{\Pr}(T=1\mid \boldsymbol X=\boldsymbol x , \boldsymbol C=\boldsymbol c, \boldsymbol W=\boldsymbol w )$ for all tested subjects.

A proof of the consistency of this estimator under the independence assumption~(\ref{indep}) is given in the eAppendix Section 3. It is required that the models for ${Q}_{Y^1,T=1}(\boldsymbol x, \boldsymbol c, \boldsymbol w)$ and ${\Pr}(T=1\mid \boldsymbol X=\boldsymbol x , \boldsymbol C=\boldsymbol c, \boldsymbol W=\boldsymbol w )$ are both correctly specified. 
Additionally, in the eAppendix Section 1.2 we present an extension of this estimator that can accommodate imperfect test sensitivity and specificity.

\section{Simulation Study}\label{sect_simulation}

In order to evaluate the proposed IPW method under the mDAG in Figure~\ref{DAG1}, and compare it to the proper TND analysis in terms of the parameter of interest and performance, we performed a simulation study. We evaluated IPW under both correctly and incorrectly specified models, with both measured and unmeasured symptom status. We also implemented a setting where a variable can impact both infection and tested status. 

We simulated ordered data $\boldsymbol O^{*}=(C,X,Y^1,W,T)$, where each variable is unidimensional, for a population of 1,000,000, and then i) randomly sampled 2000 tested participants and 2000 members of the general population and ii) randomly sampled 2000 tested \emph{symptomatic} participants (the proper TND), respectively. Full details are in the eAppendix Section 4. We established the true conditional ORs between $X$ and $Y^1$ in the general population ($\exp(\beta)$) and among those with symptoms ($\exp(\beta^{*})$, the ``relative'' OR), respectively.

In order to contrast the different quantities being estimated and the performance of different approaches, we ran three logistic regression analyses to measure the association between X and $Y^1$ adjusted for $C$. The first corresponds to the analysis in the proper TND which selects patients with $T=1$ and $W=1$. The second compares risk factors $X$ between population controls and symptomatic test-positives in the proper TND.~\cite{vdb2020} The third subsets on those tested ($T=1$) among those symptomatic and asymptomatic in order to demonstrate the collider bias arising in the absence of any kind of adjustment for $W$. We also ran three implementations of IPW with: a) correctly specified logistic regression models for $Q_{Y,T=1}(x,c,w)$ and $\Pr(T=1\mid X=x,C=c,W=w)$ where the latter regression is weighted using the case-control weights; b) a misspecified testing model which omits an interaction between $W$ and $T$; and c) no adjustment for $W$. In these IPW analyses, we set the assumed testing prevalence $\hat{q}_0$ to the truth (roughly 0.2\% of the full population depending on the simulated data). The R code for the IPW estimator is given in the eAppendix Section 5.

We evaluated these estimators under three scenarios: 1) $X$ increases the risk of any infection but more so for SARS-CoV-2 such that $\exp(\beta)>1$ and $\exp(\beta^*)>1$; 2) $X$ increases the risk of infection for SARS-CoV-2 to the same extent as for other infection on the OR scale ($\exp(\beta)>1$ and $\exp(\beta^*)$ is close to one); and 3) same as scenario 1 but we added a variable, unrelated to $(X,C)$, that affects both SARS-CoV-2 infection and testing. Scenario 2 relates to the car example from Section~\ref{sect_collider}. The additional variable in Scenario 3 reflects a violation of  the assumptions underlying the proper TND analysis. An example of such a variable may be health-care-seeking behaviour (HCSB)~\cite{Sullivan}.


To construct 95\% confidence intervals for the IPW method, we used a case-control nonparametric bootstrap method, where resampling with replacement is done separately in the tested and untested groups.~\cite{wang_ccbs} 
All simulations were run with R statistical software v. 3.6.1~\cite{R}.

The results of all implementations in addition to the analysis conducted only on tested subjects are given in Table~\ref{results}. The true values of the OR parameters (prospective and relative) are given for each scenario. We note that in the second scenario, where the risk factor $X$ has an equal strength for both types of infection, the true value of the relative OR is greater than one. Mean parameter estimates, Monte Carlo standard errors, and $\%$ coverage of the 95\% confidence intervals are given, the latter only for the relevant parameter. In the first two scenarios, the analysis of the proper TND was unbiased for the relative OR and the analysis of the symptomatic test-positives versus controls produced unbiased estimates for the true prospective OR. But both were biased in the presence of HCSB. In the design that recruits tested individuals ($T=1$), the logistic regression analysis was highly biased.
IPW implemented with correct models had no error on average, with 95\% bootstrap confidence intervals producing slight undercoverage.  IPW was  biased when the model for testing was missing an interaction term or confounder or when $W$ or HCSB was considered to be unmeasured, where bias was on the same order as the uncorrected analysis of only tested subjects. 

\begin{table}[]
\centering
\begin{tabular}[t]{lllll}
\toprule
&\textbf{Mean est}& \textbf{MC SE} &  \textbf{\% Cov $\beta$} & \textbf{\% Cov $\beta^*$}\\
\midrule
\multicolumn{5}{l}{\emph{Scenario 1}: $X$ is a risk factor for other infection, and a stronger risk factor for SARS-CoV-2}\\
\\

\textbf{True OR $\exp(\beta)$ = 2.5  }   &&& &\\
\textbf{True relative OR $\exp(\beta^*)$ = 1.95 }&  & & &\\
                &                   &                    &       &                      \\
Analysis of proper TND ($T=1$ and $W=1$)                                 &  1.94  &0.16&-&96   \\

Analysis of symptomatic test+ vs controls                               &2.44&0.13&95& -  \\
Analysis of all tested subjects ($T=1$)                                           &  1.60  &0.18   &26     & -      \\
IPW                 &                    &       &             &     
\\
\quad Correct models          & 2.52     & 0.23  &  91     &-  \\
\quad Missing interaction      & 1.85 &0.23&66&-       \\
\quad Omitted $W$       & 1.60  &0.18&24& -     \\

\midrule

\multicolumn{5}{l}{\emph{Scenario 2}: Car example: $X$ is risk factor of same strength for SARS-CoV-2 and other infection}\\
\\

\textbf{True OR $\exp(\beta)$ = 1.5  }   &&& &\\
\textbf{True relative OR $\exp(\beta^*)$ = 1.16 }&  & & &\\
                &                   &                    &       &                      \\
Analysis of proper TND ($T=1$ and $W=1$)                                 &  1.17  &0.15&-& 94  \\

Analysis of symptomatic test+ vs controls     &1.48&0.15&94& -  \\
Analysis of all tested subjects ($T=1$)                                           &        0.96         &  0.20   &  37    &    -   \\
IPW                 &                    &       &             &     
\\
\quad Correct models          & 1.50   & 0.27  & 90      &-  \\
\quad Missing interaction      & 1.09 &0.28&69&    -   \\
\quad Omitted $W$       & 0.96  &0.20&31&    -  \\

\midrule

\multicolumn{5}{l}{\emph{Scenario 3}:  Health-care-seeking behaviour affects SARS-CoV-2 and tested status}\\
\\

\textbf{True OR $\exp(\beta)$ = 3.11   }   &&& &\\
\textbf{True relative OR $\exp(\beta^*)$ = 2.45 }&  & & &\\
                &                   &                    &       &                      \\
Analysis of proper TND ($T=1$ and $W=1$)                                 &  3.83  &0.12&-&3   \\

Analysis of symptomatic test+ vs controls     &5.75&0.11&0& -  \\
Analysis of all tested subjects ($T=1$)                                           &  4.81               & 0.19    &   38   &  -     \\
IPW, omitting HCSB variable       &               4.87     &   0.29    &     61        &     -
\\
IPW, adjusting for HCSB variable       &   3.13         &  0.43  &    92     &     -
\\

\bottomrule
\end{tabular}\caption{Aggregate results of the application of each method and implementation on 1000 simulated datasets of 2000 untested population controls and 2000 tested individuals. Mean est: exponential of the mean estimate of $\beta$ (i.e. transformed to the OR scale); MC SE: Monte-Carlo standard error of $\hat\beta$; \% Cov: \% of 95\% confidence intervals that contain the true $\beta$ or $\beta^*$ (optimal is 95\%); HCSB: health-care-seeking behaviour. 
\label{results}}
\end{table}

We present the results of an additional simulation study in the eAppendix Section 6 that investigates the robustness of the IPW estimator under incorrect values of test sensitivity and specificity and incorrect knowledge of the prevalence of testing.

\section{Discussion}

In this paper, we have contributed to the investigation of statistical analysis under designs that recruit patients at test sites in the context of evaluating risk or preventive factors of SARS-CoV-2. We defined a potential parameter of interest in such a study as the coefficients in a regression model for the true infection outcome. We formally differentiated between this parameter and the parameter targeted in a proper TND. We explained and demonstrated the importance of sampling additional population controls~\cite{vdb2020} with the goal of identifying risk factors that increase the risk of infection with SARS-CoV-2 specifically. We then investigated 
the advantages and disadvantages of undertaking a study design that recruits among all people (symptomatic and asymptomatic) being tested for SARS-CoV-2 as opposed to only symptomatic patients (as in the proper TND) in addition to population controls. Finally, we proposed a novel IPW estimator, which does not rely on parametric modeling assumptions~\cite{Pirracchio}, that can be employed under the former design to statistically adjust for collider bias under additional data collection. We then evaluated different estimation approaches in several settings through simulation study.

Recent work by Shi et al evaluated the identifiability of vaccine effectiveness parameters under case-control designs and TNDs.~\cite{Shi2017} Foppa et al derived theoretical results, supported through simulation study, regarding the interpretability of vaccine effectiveness parameters and potential sources of selection bias under the TND. These results are important as they shed light on the data collection needed to correctly estimate a parameter of interest under a given study design. In the setting where all tested subjects are recruited, our results indicated that we must measure all variables on the pathway between infection and testing in addition to all common causes of infection and testing. This means that incomplete ascertainment of the symptoms or other reasons leading some individuals to be tested may result in a biased estimator. The survey can be designed to accommodate this, through open questions relating to reasons for seeking testing. 
We must also measure and adjust for all causes of testing if they are also causes of SARS-CoV-2 infection and/or symptoms. 
We also formally assumed independence between the study participants and suggested that limiting recruitment to one member per household could help support this assumption. But in general the interpretation of risk factor associations with ``contagious outcomes'' is complicated by differential exposure to disease.~\cite{Morozova2018}

Due to the rapidly evolving nature of the COVID-19 pandemic, studies with short timelines are necessary to monitor public health. The accessibility of a design that recruits participants at testing sites allows for much faster results compared to a cohort study of initially uninfected individuals. But adjustment for bias must occur either through study design or statistical control for selection factors leading to testing. We stress the importance of defining the parameters of interest in a risk factor analysis, investigations of the identifiability of these parameters, and the development of estimators that are consistent under assumptions about the data-generating mechanism and design. These steps allow for a principled approach that may help avoid substantial sources of bias when tracking risk and preventive factors of SARS-Cov-2 infection. 

\bibliographystyle{ama.bst}
\bibliography{references}

\end{document}